\documentclass[11pt,a4paper]{scrartcl}

\usepackage[utf8]{inputenc}
\usepackage[english]{babel}           
\usepackage{amsfonts, amsmath, amsthm, amssymb, dsfont}
\usepackage{graphicx}
\usepackage{epstopdf}
\usepackage{color}

\usepackage{lineno}

\usepackage{natbib}

\newcommand{\s}{\footnotesize}  
\newcommand{\xs}{\scriptsize} 

\title{Modulational stability of nonlinear saturated gravity waves}
\author{Mark Schlutow}
\date{}

\begin{document}

	\maketitle

	\begin{abstract}
		Stationary gravity waves, such as mountain lee waves, are effectively described by Grimshaw's dissipative modulation equations even in high altitudes where they become nonlinear due to their large amplitudes. 
In this theoretical study, a wave-Reynolds number is introduced to characterize general solutions to these modulation equations. This non-dimensional number relates the vertical linear group velocity with wavenumber, pressure scale height and kinematic molecular/eddy viscosity. It is demonstrated by analytic and numerical methods that Lindzen-type waves in the saturation region, i.e. where the wave-Reynolds number is of order unity, destabilize by transient perturbations. It is proposed that this mechanism may be a generator for secondary waves due to direct wave-mean-flow interaction. By assumption the primary waves are exactly such that altitudinal amplitude growth and viscous damping are balanced and by that the amplitude is maximized. 
Implications of these results on the relation between mean-flow acceleration and wave breaking heights are discussed.
	\end{abstract}

	\section{Introduction}


%
	Atmospheric gravity waves generated in the lee of mountains extend over scales across 
	which the background may change significantly.
	The wave field can persist throughout the layers from the troposphere to the deep atmosphere, 
	the mesosphere and beyond \citep{Fritts2016,Fritts2018}.
	On this range background temperature and therefore stratification and background density
	may undergo several orders of magnitude in variation.
	Also, dynamic viscosity and thermal conductivity cannot be considered constant on such a domain \citep{Pitteway1963,Zhou1995}.

	Two predominant regimes for the waves can be identified: the homosphere and the heterosphere.
	These two are separated by the turbopause, usually somewhere in the mesopause region. 
	Below the turbopause molecular viscosity is negligible.
	Hence, if diffusion occurs, it is caused by turbulence. 
	Due to the missing damping and the thinning background air, mountain waves amplify exponentially when extending upwards.
	This phenomenon is also called anelastic amplification.

	At certain heights the amplitudes cannot grow any further due to limiting processes.
	Those may be static or dynamic instabilities that act on the small scale comparable to the wavelength.
	For instance, \citet{Klostermeyer1991} showed that all inviscid nonlinear Boussinesq waves are prone to parametric instabilities.
	The waves do not immediately disappear by the small-scale instabilities, rather the perturbations grow comparably slowly
	such that the waves persist in their overall structure over several more wavelengths.
	However, turbulence is produced.
	\citet{Lindzen1970} modeled the effect of turbulence on the wave by 
	harmonic damping with a constant kinematic eddy viscosity.
	The eddy viscosity is exactly such that it saturates the wave, 
	meaning that viscous damping and anelastic amplification are balanced \citep{Lindzen1981,Fritts1984,Dunkerton1989,Becker2012}.
	\citet{Pitteway1963} referred to this instance as amplitude-balanced wave.

	Above the turbopause molecular viscosity takes over. It is modeled by a constant dynamic viscosity.
	In combination with the thinning background density, the kinematic viscosity increases exponentially 
	with height resulting in a rapid decrease in amplitude.

	In the process of becoming saturated the amplitude becomes considerably large such that the waves
	cannot be considered linear.
	Pioneering work on the mathematical description of nonlinear gravity waves was accomplished by \citet{Bretherton1966} and \citet{Grimshaw1972}.	
	This paper aims to extent Lindzen's linear wave saturation theory with the aid of Grimshaw's nonlinear wave description.

\section{How the modulation equations solve the compressible Navier-Stokes equations -- a brief review}

	The nonlinear governing equations for our investigations are 
	the two-dimensional dissipative Grimshaw's modulation equations being first introduced by
	\citet{Grimshaw1974}.
	Before presenting them, we review in this section how they solve the dimensionless compressible, 
	Reynolds-averaged Navier-Stokes equations (NSE) asymptotically.
	Length and time are nondimensionalized via 
	\begin{align}
		(x^\ast,z^\ast,t^\ast)=\left(L_\mathrm{r}\hat x,\,L_\mathrm{r}\hat z,\, \frac{L_\mathrm{r}}{\upsilon_\mathrm{r}} \hat t\right)
	\end{align}		
	where $L_\mathrm{r}\approx 1\dots 10$\,km denotes the reference wavelength (hence the subscript r) and $\upsilon_\mathrm{r}$ is the reference velocity.
	To separate the hydrostatic background from the flow field associated with the wave the first ingredient necessary 
	is a small scale separation parameter 
	\begin{align}
		\varepsilon = L_\mathrm{r}/H_\theta \ll 1
	\end{align}
	where $H_\theta\approx 10\dots 100$\,km is the potential temperature scale height.
	This choice for the scale separation parameter was introduced by \citet{Achatz2010}.

	The authors considered inviscid flows. 
	In order to take viscous damping into account, we need to compare inviscid and viscous terms.
	The (eddy) viscosity is not a constant throughout the atmosphere and among others depends on temperature. 
	\citet[][Fig. 9]{Midgley1966} gave a realistic
	vertical profile of the effective kinematic viscosity combining eddy and molecular effects. 
	\citet{Hodges1969} computed the eddy diffusion 
	by gravity waves near the mesopause to be $K_\mathrm{{r(eddy)}}\approx 10^{6\dots 7}\,\mathrm{cm}^2\mathrm{s}^{-1}$
	which compares well with \citet{Midgley1966} and \citet{Lindzen1981}.
	In the scaling regime of \citet{Achatz2010} as well as \citet{Schlutow2017b} these values correspond to Reynolds numbers
	of $\mathrm{Re}\approx 10\dots 100$ when the Mach number is $\mathrm{Ma} \approx 0.1\dots 0.01$. 
	These numbers are also supported by a review of \citet{Fritts1984}.
	
	Taking these arguments into account, 
	a realistic flow regime in terms of Mach, Froude, Reynolds and Prandtl number 
	most suitable for internal gravity waves in the middle/upper atmosphere region 
	is then found by assuming 
	\begin{subequations}
	\label{eq:ndcn}
	\begin{align}
			\frac{\upsilon_\mathrm{r}}{\sqrt{(1-\kappa)p_\mathrm{r}/\rho_\mathrm{r}}}\equiv\mathrm{Ma}&=\mathit{O}(\varepsilon)\\
			\frac{\upsilon_\mathrm{r}}{\sqrt{gL_\mathrm{r}}}\equiv\mathrm{Fr}&=\mathit{O}(\varepsilon^{1/2})\\
			\frac{\rho_\mathrm{r}\upsilon_\mathrm{r}L_\mathrm{r}}{ \mu_\mathrm{r}}\equiv\mathrm{Re}&=\mathit{O}(\varepsilon^{-1})\\
			\frac{c_p \mu_\mathrm{r}}{\kappa_\mathrm{r}}\equiv\mathrm{Pr}&=\mathit{O}(1)
	\end{align}
	\end{subequations}
	where $\rho_\mathrm{r}, p_\mathrm{r}, \kappa_\mathrm{r}$ and  $\mu_\mathrm{r}$ 
	represent the reference density, pressure, thermal conductivity and dynamic viscosity, respectively.
	The constant $\kappa=2/7$ is the ratio of the ideal gas constant $R$ to the specific heat capacity at
	constant pressure $c_p$ for diatomic gases.
	Equations \eqref{eq:ndcn} provide a distinguished limit for multiple scale asymptotic analysis.
	Introducing compressed coordinates for the horizontal, vertical and time axis
	\begin{align}
		(x,\,z,\,t)=(\varepsilon\hat x,\,\varepsilon\hat z,\,\varepsilon\hat t)
	\end{align}		
	separates the slowly varying background from the fast oscillating wave fields.
	Under the assumption of stable stratification to the leading order, 
	the hydrostatic background is determined 
	by the vertical temperature profile $T(z)$.
	Two dimensionless background variables,
	that vary on the large scale,
	will appear in the final modulation equations,
	the Brunt-V\"ais\"al\"a frequency or stratification measure $N$ and the background density $\rho$.
	They have to be calculated from the temperature by solving 
	\begin{align}
		\label{eq:bg_ode1}
		N^2 &= \frac{1}{T}\left(\frac{\mathrm{d}T}{\mathrm{d}z}+1\right)\\
		\label{eq:bg_ode2}		
		\frac{1}{\rho}\frac{\mathrm{d}\rho}{\mathrm{d}z} &= -\frac{1}{T}\left(\frac{\mathrm{d}T}{\mathrm{d}z}+\frac{1}{\kappa}\right)
	\end{align}		
	that originate from the hydrostatic assumption in combination with the ideal gas equation of state.
	
	With the given scale separation of background and wave field, we can formulate the scaled two-dimensional compressible NSE,
	\begin{subequations}
    \label{eq:psinclike3}
	\begin{align}
	    \frac{\mathrm{D}\mathbf{v}}{\mathrm{D}\hat t}+\hat{\nabla} p-Nb\mathbf{e}_z&=-\varepsilon \,(Nb\hat{\nabla} p-N^2p\mathbf{e}_z)
	    +\varepsilon\Lambda\hat{\nabla}^2\mathbf{v}+\dots\\
	    \frac{\mathrm{D}b}{\mathrm{D}\hat t}+Nw&=-\varepsilon \,\left(N^2+\frac{1}{N}\frac{\mathrm{d}N}{\mathrm{d}z}\right)wb
	    +\varepsilon\Lambda\hat{\nabla}^2b+\dots\\
	   \hat{\nabla}\cdot\mathbf{v}&=-\varepsilon \,\left(N^2+\frac{1}{\rho}\frac{\mathrm{d}\rho}{\mathrm{d}z}\right)w+\dots,
	\end{align}	
	\end{subequations}
	in terms of the wave field variables, velocity $\mathbf{v}$, buoyancy $Nb$, and kinematic pressure $p$ 
	where $\hat{\nabla}=(\partial/\partial\hat{x},\,\partial/\partial\hat{z})^\mathrm{T}$ 
	and $\mathbf{e}_z$ the unit vector pointing in the vertical direction.
	$\mathrm{D}/\mathrm{D}\hat{t}$ denotes the material derivative and $\Lambda$ is the total kinematic viscosity.
	The dots represent higher-order terms that we do not give explicitly but must not be neglected.
	
	Wentzel-Kramers-Brillouin (WKB) theory provides us with a multiple scale method 
	to solve the scaled NSE asymptotically by the spectral ansatz
	\begin{align}
		\label{eq:asym_expa}
		U(\hat x,\hat z,\hat t;\varepsilon)=U_{0,0}(x,z,t)+\left(U_{0,1}(x,z,t)e^{i\phi(x,z,t)/\varepsilon}
		+\mathrm{c.c.}\right)+\mathrm{h.h.}+\mathit{O}(\varepsilon)
	\end{align}
	where c.c. stands for the complex conjugate and h.h. for higher harmonics.
	The vector $U=(\mathbf{v},\,b,\,p)^\mathrm{T}$ contains the prognostic variables and $\phi$ is the wave's phase.
	This ansatz is inserted into the compressible NSE and terms are ordered with respect to powers of $\varepsilon$
	and harmonics.

	\section{Governing equations: Grimshaw's dissipative modulation equations}
	To leading order of the nonlinear WKB analysis of the NSE with the turbulence model of Lindzen
	one obtains Grimshaw's dissipative modulation equations 
	supported by a mean-flow horizontal kinematic pressure gradient \citep[Eqs. (4.1 -- 4.6)]{Grimshaw1974}.
	Note that the leading-order WKB analysis of the dissipative pseudo-incompressible equations \citep{Durran1989}
	lead to the same modulation equations.
	We present them in slightly different notation as Grimshaw and with the prerequisite
	that the wave field is horizontally periodic ($0<k_x=\mathrm{const}$) and only modulated in the $z$-direction,
	\begin{subequations}
	\label{eq:modeq}
	\begin{align}
		\label{eq:modeq_kz}		
		\frac{\partial k_z}{\partial t}+\frac{\partial \omega}{\partial z}&=0\\
		\label{eq:modeq_a}	
		\rho\frac{\partial a}{\partial t}+\frac{\partial \omega'\rho a}{\partial z}&=-\Lambda |\mathbf{k}|^2\rho a\\
		\label{eq:modeq_u}	
		\rho\frac{\partial u}{\partial t}+\frac{\partial \omega'k_x\rho a}{\partial z}&=-\frac{\partial p}{\partial x}.
	\end{align}
	The modulation equations govern the evolution of vertical wavenumber $k_z$, wave action density $\rho a$, and mean-flow horizontal wind $u$.
	Equations \eqref{eq:modeq_kz}--\eqref{eq:modeq_u} are closed by
	\begin{align}
		|\mathbf{k}|^2&=k_x^2+k_z^2\quad\\
		\label{eq:dispersion}
		\omega&=\frac{Nk_x}{|\mathbf{k}|}+k_xu\\
		\omega'&=-\frac{Nk_xk_z}{|\mathbf{k}|^3}
	\end{align}
	\end{subequations}
	where $\mathbf{k}$, $\omega$, and $\omega'$  represent the wavenumber vector, extrinsic frequency and 
	vertical linear group velocity, respectively.
	Note that primes denote derivative with respect to the vertical wavenumber throughout this paper.
	Extrinsic frequency is defined by the sum of intrinsic frequency and Doppler shift.
	The prognostic variables determine the asymptotic solution
	as described in the previous section and explained in \citet{Schlutow2017b}.
		
	The first equation \eqref{eq:modeq_kz} describes the evolution of the phase in \eqref{eq:asym_expa} 
	as by definition $\nabla \phi\equiv\mathbf{k}$ and $-\partial\phi/\partial t\equiv\omega$.		
	The second equation \eqref{eq:modeq_a} governs the conservation of wave action density 
	being the ratio of wave energy density and the intrinsic frequency. 
	In terms of the polarization relation, the leading-order first harmonics $U_{0,1}$ are computable from the wave action density.
	The	third equation \eqref{eq:modeq_u} accounts for the acceleration of the horizontal mean-flow wind $u_{0,0}$.
	The variable $p$ corresponding to the mean-flow horizontal kinematic pressure $p_{0,0}$ 
	is unknown and needs additional investigation to close the system.
	Note that we dropped the indeces for the mean-flow variables.

	The governing equations \eqref{eq:modeq} can be reformulated in vector form
	\begin{align}
		\label{eq:modeq_vec}
		\frac{\partial \mathbf{y}}{\partial t}+\frac{\partial \mathbf{F}(\mathbf{y})}{\partial z}=\mathbf{G}(\mathbf{y})
	\end{align}
	with a flux $\mathbf{F}$ and an inhomogeneity $\mathbf{G}$
	where $\mathbf{y}=(k_z, a, u)^\mathrm{T}$ is the prognostic vector.
	
	\section{General stationary solutions of the modulation equations}

	In this section we will explore general stationary solutions 
	before we focus on particular solutions for which we will present stability analysis in the upcoming sections.
	
	In the inviscid limit, i.e. $\Lambda\rightarrow 0$, the modulation equations assume stationary solutions
	where ${\partial p/\partial x=0}$ which can be computed analytically by a formula of \citet[][Eq. 5.20]{Schlutow2017b}.
	When we multiply \eqref{eq:modeq_a} by $k_x$ and subtract \eqref{eq:modeq_u},
	we obtain
	\begin{align}
		\label{eq:stat_cond}
		\rho\frac{\partial }{\partial t}(k_xa-u)=\frac{\partial p}{\partial x}-k_x\Lambda |\mathbf{k}|^2\rho a
	\end{align}
	Thus, to be coherent with the inviscid limit, the dissipative modulation equations assume stationary solutions
	only if the right hand side of \eqref{eq:stat_cond} vanishes which provides eventually a closure 
	for the mean-flow horizontal kinematic pressure gradient,
	\begin{align}
		\label{eq:mfpg}
		\frac{\partial p}{\partial x}=k_x\Lambda |\mathbf{k}|^2\rho a.
	\end{align}
	This result implicates that the mean-flow horizontal kinematic pressure gradient balances the viscous forces acting on the horizontal mean-flow wind.
	
	Then, a general stationary solution $\mathbf{Y}(z)=(K_z,\,A,\,U)^\mathrm{T}$ depicting a mountain lee wave fulfills
	\begin{align}
		\label{eq:stat_dispersion}
		\omega(K_z,U) &=0,\\
		\frac{\partial\Omega'\rho A}{\partial z}&=-\Lambda |\mathbf{K}|^2\rho A,\\
		U&=U(z)
	\end{align}
	with $|\mathbf{K}|^2=K_x^2+K_z^2$ and $\Omega'=\omega'(K_z)$.
	Note that we label the stationary solution by capital letters.
	Given any reasonable mean-flow horizontal wind $U(z)$, 
	the remaining two variables of the general stationary solution are given explicitly by
	\begin{align}
		\label{eq:statio_kz}
		K_z&=-\sqrt{\frac{N^2}{U^2}-K_x^2},\\
		\label{eq:statio_a}
		A&=\frac{\mathcal{F}}{\rho\Omega'}\exp\left(-\int_0^z\frac{\Lambda |\mathbf{K}|^2}{\Omega'}~\mathrm{d}\tilde z\right).
	\end{align}
	with $\mathcal{F}$ an integration constant.
	
	\section{Lindzen-type mountain lee wave and the wave-Reynolds number}	

	So far, we considered all background variables of our governing PDE as functions of $z$.
	In this section we will prescribe these functions in a piecewise fashion 
	in order to construct a typical mountain lee wave that gets saturated 
	by some small-scale instability process comparable to \citet{Lindzen1981}.
	The complete solution is divided into an unsaturated as well as a saturated middle atmospheric part
	and a deep atmosphere part.

	\subsection{The unsaturated middle atmospheric solution}	

	First, we assume that the background atmosphere is piecewise isothermal.
	From \eqref{eq:bg_ode1}, this assumption implies that $N=\mathrm{const}$, some typical value for the middle atmosphere.
	Then, \eqref{eq:bg_ode2} gives 
	\begin{align}
		\rho(z)=\rho_0e^{-z/H}
	\end{align}		
	where $H=\kappa N^{-2}=\mathrm{const}$ denotes the dimensionless (local) pressure scale height.
	Second, we assume that the mean-flow horizontal wind is piecewise constant as well, so $U=\mathrm{const}$.
	These assumptions can be weakened which we will discuss in the concluding Sec.~\ref{sec:conclusion}.
	It follows immediately by \eqref{eq:statio_kz} and horizontal periodicity that $K_z=\mathrm{const}$ making it a plane wave.
	
	In the middle atmosphere viscosity is negligible. Therefore, below the breaking height $z_\mathrm{break}$ the integral in \eqref{eq:statio_a} vanishes
	and the amplitude of the wave grows exponentially with the inverse density,
 	\begin{align}
		A(z)&=\frac{A_0}{\rho(z)}\quad\text{in }[0,z_\mathrm{break}).
	\end{align}
	Because of the vanishing dissipation, the mean-flow horizontal kinematic pressure gradient also vanishes according to \eqref{eq:mfpg}.

	\subsection{The saturated middle atmospheric solution}
	
	Above $z_\mathrm{break}$, the wave saturates by some small-scale instability process producing turbulence
	which balances the anelastic amplification.
	The exact kinematic eddy viscosity that keeps the amplitude leveled in \eqref{eq:statio_a}, such that $A=\mathrm{const}$ , is then given by 
	\begin{align}
		\label{eq:bal_con}
		\Lambda=\Omega'H^{-1}|\mathbf{K}|^{-2}=\mathrm{const}\quad\text{in }[z_\mathrm{break},z_\mathrm{turbo}).
	\end{align}
	In this region, the mean-flow horizontal kinematic pressure gradient depends on $z$ only by the background density 
	and can be computed inserting \eqref{eq:bal_con} into \eqref{eq:mfpg},
	\begin{align}
		\frac{\partial P}{\partial x}=K_xH^{-1}\Omega'A\rho\quad\text{in }[z_\mathrm{break},z_\mathrm{turbo}).
	\end{align}
	In particular, it is positive while the mean-flow horizontal wind is negative 
	which can be seen from \eqref{eq:dispersion} and \eqref{eq:stat_dispersion}.

	\subsection{The deep atmosphere solution}
	
	The saturated middle atmospheric solution is valid below the turbopause at $z_\mathrm{turbo}$ 
	as in the heterosphere the molecular viscosity dominates 
	which is modeled by a constant dynamic viscosity $\mu_{mol}$ implying
	\begin{align}
		\Lambda(z)=\frac{\mu_{mol}}{\rho(z)}\quad\text{in }[z_\mathrm{turbo},+\infty).
	\end{align}
	In the deep atmospheric region the integral in \eqref{eq:statio_a} has also an analytic solution for $A$
	which decays quickly for $z\rightarrow+\infty$.
	Here, typical values $N$ and $U$ for the deep atmosphere are used.

	\subsection{The wave-Reynolds number}
	
\begin{figure*}
	\begin{center}
		\input{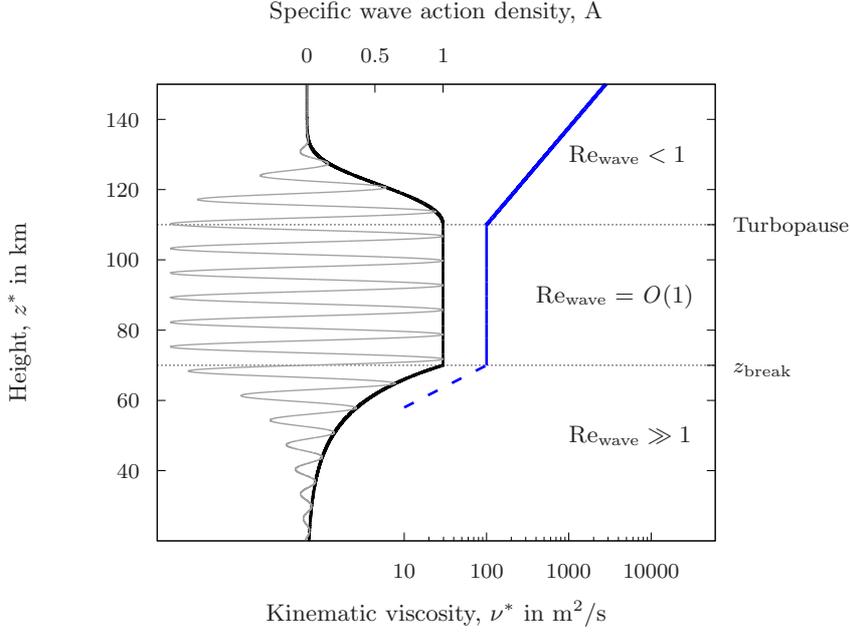}	
	\end{center}
	\caption{Schematic of a saturated plane mountain lee wave (thin grey line) 
		with amplitude profile (thick black line) 
		and effective kinematic viscosity (thick blue line)
		being characterized by the wave-Reynolds number.}
	\label{fig:sat_wave}
	\end{figure*}

	In Fig.~\ref{fig:sat_wave}, such a prototypical mountain wave, 
	that saturates, is illustrated. 
	By inspection of equation \eqref{eq:bal_con}, its behavior in the different regions may be characterized by a ``wave-Reynolds'' number.
	It can be written as
 	\begin{align}
		\frac{\Omega'H^{-1}|\mathbf{K}|^{-2}}{\Lambda }\gtreqqless 1.
	\end{align}
	When reintroducing the dimensional variables by reversing the non-dimensionalization of \citet{Schlutow2017b}, so
	\begin{align}
		H=\frac{\varepsilon}{L_\mathrm{r}}H_p^\ast,\quad
		\Lambda=\frac{\rho_\mathrm{r}}{\mu_\mathrm{r}}\nu^\ast,\quad
		N=\frac{L_\mathrm{r}}{\upsilon_\mathrm{r}} N^\ast,\quad
		\mathbf{K}=L_\mathrm{r}\mathbf{K}^\ast,
	\end{align}
	and using the definitions \eqref{eq:ndcn},
	the wave-Reynolds number reads
	\begin{align}
		\mathrm{Re}_\mathrm{wave}\equiv\frac{C_{gz}D}{\nu^\ast}.
	\end{align}
	The dimensional linear vertical group velocity is denoted by $C_{gz}$ and represents the velocity scale for this type of Reynolds number.
	The term $D=H_p^{\ast-1}|\mathbf{K}|^{\ast-2}$ defines its length scale.
	Estimates of the wave-Reynolds number in the different regions are depicted in Fig.~\ref{fig:sat_wave}.

	\section{Stability of the saturated wave}
	\label{sec:stability}

	In this section we will investigate the saturated mountain wave with respect to stability.
	In particular, we are interested in the region where $\mathrm{Re}_\mathrm{wave}=\mathit{O}(1)$
	as here the amplitude is at its maximum and one can expect most likely nonlinear behavior.
	We assess stability by analyzing the evolution of small perturbations.
	Due to their smallness, we can linearize the governing equations in vector form \eqref{eq:modeq_vec} 
	around the stationary solution $\mathbf{Y}$.
	Applying the ansatz
	\begin{align}
		\mathbf{y}(z,t)=\hat{\mathbf{y}}(z)e^{\lambda t}
	\end{align}	 
	for the perturbation, results in an eigenvalue problem (EVP)
	\begin{align}
		\label{eq:evp}
		\lambda \mathbf{y}+\frac{\partial }{\partial z}\left(\left.\frac{\partial \mathbf{F}}{\partial \mathbf{y}}\right|_\mathbf{Y}\,\mathbf{y}\right)
		=\left.\frac{\partial \mathbf{G}}{\partial \mathbf{y}}\right|_\mathbf{Y}\,\mathbf{y}
	\end{align}
	where we dropped the hat over $\mathbf{y}$.
	The Jacobian matrices of the flux and inhomogeneity evaluated at $\mathbf{Y}$ are given by
	\begin{align}
		\left.\frac{\partial \mathbf{F}}{\partial \mathbf{y}}\right|_\mathbf{Y}&=
		\begin{pmatrix}
			\Omega'& 0& K_x\\
			\Omega''A& \Omega'& 0\\
			K_x\Omega''A& K_x\Omega'& 0
		\end{pmatrix},\\
		\left.\frac{\partial \mathbf{G}}{\partial \mathbf{y}}\right|_\mathbf{Y}&=
		\begin{pmatrix}
			0& 0& 0\\
			H^{-1} \Omega''A-2\Lambda K_zA& 0& 0\\
			K_xH^{-1}\Omega''A-2K_x\Lambda K_zA& 0& 0
		\end{pmatrix}.
	\end{align}

	Note that the second and third row of each Jacobian are linearly dependent.
	Hence, we can solve the third equation of \eqref{eq:evp} for 
	\begin{align}
		u=K_xa
	\end{align}
	which reduces the dimension of the system, 
	so with a slight abuse of notation, ${\mathbf{y}=(k_z,a)^\mathrm{T}}$ and 
	\begin{align}
		\left.\frac{\partial \mathbf{F}}{\partial \mathbf{y}}\right|_\mathbf{Y}&=
		\begin{pmatrix}
			\Omega'& K_x^2\\
			\Omega''A& \Omega'
		\end{pmatrix}\\
		\left.\frac{\partial \mathbf{G}}{\partial \mathbf{y}}\right|_\mathbf{Y}&=
		\begin{pmatrix}
			0& 0\\
			H^{-1} \Omega''A-2\Lambda K_zA& 0
		\end{pmatrix}.
%
%
	\end{align}
	If one assumes that the perturbation decays sufficiently fast towards the edges of the region 
	where $\mathrm{Re}_\mathrm{wave}=\mathit{O}(1)$ (in Fig.~\ref{fig:sat_wave} at 70 and 110\,km),
	in other words that the edges have no influence on the perturbation, 
	then one can extent the domain to the infinities.
	Consequently, we consider the differential operator associated with the EVP as closed and densely defined on $L^2$,
	the space of vector valued square integrable functions on the real line equipped with the norm
	\begin{align}
		\|\mathbf{y}\|_{L^2}=\sqrt{\int_{-\infty}^{+\infty}k_z^2+a^2+u^2~\mathrm{d}z}.
	\end{align}
	In conclusion, we translated the problem of stability	to finding the spectrum of a linear operator of an EVP.
	Broadly speaking, the spectrum is the set of all $\lambda\in\mathbb{C}$ for which the EVP has solutions.

	The EVP is solved by the Fourier transform
	\begin{align}
		\mathbf{y}=\int_{-\infty}^{+\infty}\tilde{\mathbf{y}}e^{i\mu z}~\mathrm{d}\mu
	\end{align}
	which yields an algebraic equation
	\begin{align}
		\left(\lambda+\left.\frac{\partial \mathbf{F}}{\partial \mathbf{y}}\right|_\mathbf{Y}i\mu
		-\left.\frac{\partial \mathbf{G}}{\partial \mathbf{y}}\right|_\mathbf{Y}\right)\tilde{\mathbf{y}}=0.
	\end{align}
	It has nontrivial solutions only if the coefficient matrix is singular
	which happens if
	\begin{align}
		(\lambda+\Omega'i\mu)^2
		-K_x^2A\bigl[(2\Lambda K_z-H^{-1}\Omega'')i\mu-\Omega''\mu^2\bigr]=0.
	\end{align}	 
	This characteristic polynomial has two zeros
	\begin{align}
		\label{eq:spectrum}
		\lambda_{1,2}(\mu)&=-\Omega'i\mu\pm K_x\sqrt{A}\sqrt{(
		2\Lambda K_z-H^{-1}\Omega'')i\mu-\Omega''\mu^2}
	\end{align}
	which determine the spectrum of the linear operator of the EVP as curves in the complex plane parametrized  by $\mu$,
	the spatial eigenvalue.
	$\mu$ can also be interpreted as the vertical wavenumber of the perturbation.
	The real part of $\lambda$ represents the instability growth rate,
	if it is positive,	and the imaginary part is a frequency.
	Thus, \eqref{eq:spectrum} provides also a dispersion relation
	linking the perturbation's wavenumber with its frequency. 

	One can readily show that either $\lambda_{1}$ or $\lambda_{2}$ of \eqref{eq:spectrum} has positive real part 
	for reasonable wave solutions implying that they are unconditionally unstable.
	We want to point out that when $H\rightarrow\infty$ and $\Lambda\rightarrow 0$, \eqref{eq:spectrum} reduces to the spectrum
	of the inviscid nonlinear Boussinesq plane waves \citep{Schlutow2018} 
	having the classical modulational stability criterion $\Omega''>0$ \citep{Grimshaw1977}.

	\subsection{Transient (in)stability}
	In this subsection we want to investigate the characteristics of the instability that is presented in the previous section.
	We put the ``in'' of the section title into parentheses because there is the possibility
	that the primary wave ``survives'' the instability.
	One particular type of those harmless instabilities is called transient instability. 
	Despite the fact that its norm grows exponentially in time,
	it vanishes at each given point.
	
	To show such characteristics a prerequisite for mathematical rigor must be fulfilled: 
	the linear differential operator of the EVP must be well-posed,
	which means its spectrum has a maximum real part
	or, in other words, the instability growth rate is bounded.
	The reader finds this tedious verification in the Appendix.
	We want to give an interesting remark at this point. The well-posedness depends on a criterion, $\Omega''>0$,
	which turns out to be equivalent to the modulational stability criterion 
	of plane waves in Boussinesq theory.	
	
\begin{figure*}
		\begin{center}
			\input{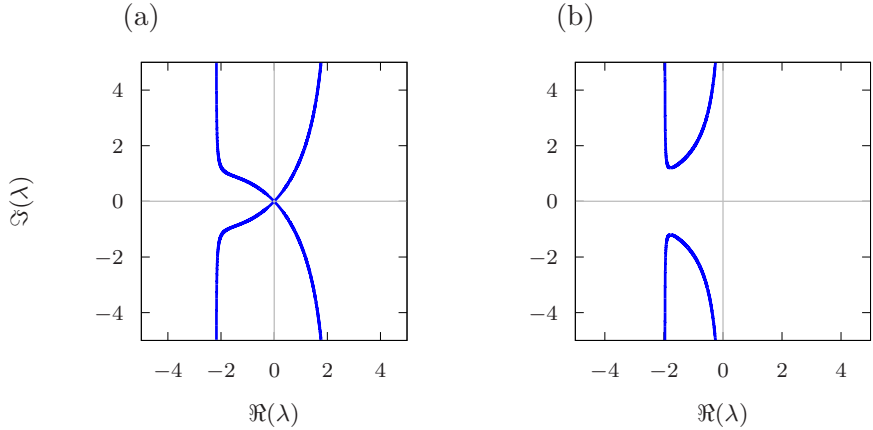}	
		\end{center}
		\caption{Unstable spectrum (blue lines)
		in $L^2$ (panel a) and stable spectrum in the exponentially weighted $L^2_\alpha$ (panel b) for the saturated wave  of Fig.~\ref{fig:pertevo}.
		The exponential weight is the largest $\alpha$ as defined by \eqref{eq:alpha}.}
		\label{fig:spectrum}
	\end{figure*}

	The key idea for transient instabilities \citep{Kapitula2013} is to ask for solutions 
	of the EVP \eqref{eq:evp} in a weighted space $L^2_\alpha$ with $\alpha\in\mathbb{R}$ an exponential weight, such that
	\begin{align}
		\|\mathbf{y}\|_{L_\alpha^2}=\|\mathbf{y}e^{\alpha z}\|_{L^2}.
	\end{align}
	Doing so, we get the same curves as in \eqref{eq:spectrum} but with $i\mu\rightarrow i\mu-\alpha$, so
	\begin{align}
		\label{eq:weighted_spectrum}
		\lambda_{1,2}^\alpha(\mu)&=-\Omega'(i\mu-\alpha)\pm K_x\sqrt{A}\sqrt{(
		2\Lambda K_z-H^{-1}\Omega'')(i\mu-\alpha)+\Omega''(i\mu-\alpha)^2}.
	\end{align}
	When we choose 
	\begin{align}
		\label{eq:alpha}
		\alpha<\frac{1}{2}\frac{2\Lambda K_z-H^{-1}\Omega''}{\sqrt{\Omega''}\Omega'+\Omega''}<0
	\end{align}
	provided $\Omega''>0$
	the spectrum is stabilized, so $\Re(\lambda_{1,2}^\alpha)\leq 0$ for all $\mu\in\mathbb{R}$.
	The negative exponential weight penalizes solutions at $-\infty$.
	To converge in the weighted norm, the perturbation in $L^2$ must therefore decay exponentially at $-\infty$ for all times. 
	But simultaneously, its $L^2$-norm grows exponentially in time.
	This seeming paradox resolves when the perturbation propagates sufficiently fast towards $+\infty$, i.e. upwards.
	Perturbations of this behavior are called transient instabilities \citep{Sandstede2000}.
	For illustrative purpose, the unweighted and weighted spectrum of an example wave, 
	that we will discuss in more detail in the following section, 
	are plotted in Fig.~\ref{fig:spectrum}.


	\subsection{Numerical investigation of the transient instabilities}
	\label{sec:num}

\begin{figure*}
		\input{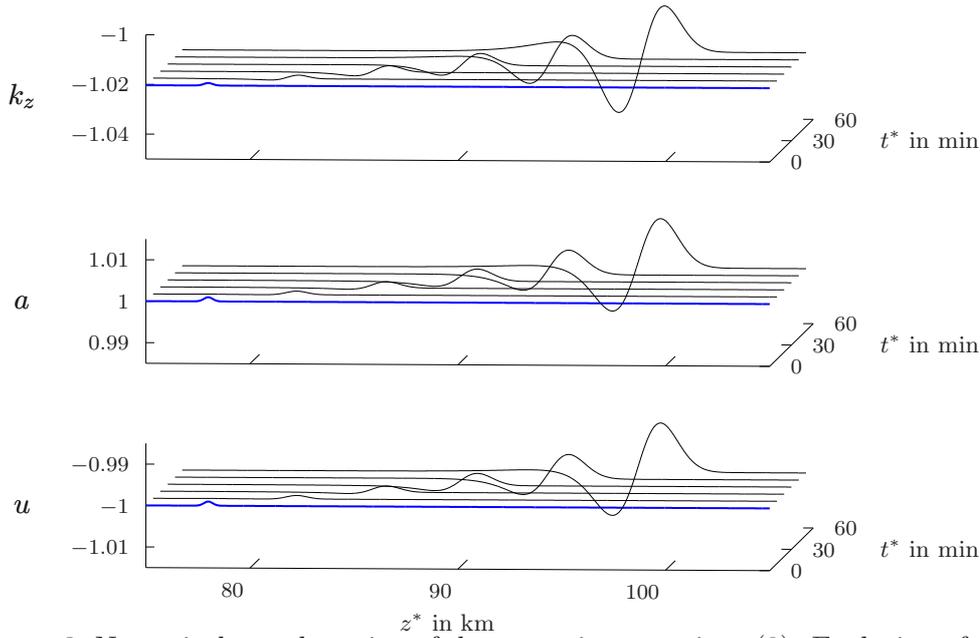}
		\caption{Numerical corroboration of the governing equations \eqref{eq:modeq}.
		Evolution of an initial Gaussian perturbation of the saturated wave centered at 78\,km (blue lines). 
		}
		\label{fig:pertevo}
	\end{figure*}

	We use the finite-volume numerical solver presented in \citet{Schlutow2018} for the governing equations \eqref{eq:modeq}
	to compute the evolution of a tiny Gaussian initial perturbation of the saturated wave in the region where ${\mathrm{Re}_\mathrm{wave}=\mathit{O}(1)}$. 
	The results are shown in Fig.~\ref{fig:pertevo}.
	The simulation is set up by $N=1$ and $K_x=1$.
	We discretize the equations on 2000 grid points and integrate over 6000 time steps.
	As can be seen in the figure, the perturbation amplifies exponentially and propagates to the right, i.e. upwards, as theory suggests.
	We can also observe that the perturbation's wavelength compares with the scale height.
	Or in other words the instability varies on the large scale.
	The corresponding spectra to this case as computed by \eqref{eq:spectrum} 
	and \eqref{eq:weighted_spectrum} are plotted in Fig.~\ref{fig:spectrum}.
	The amplitude and the vertical wavenumber undergo strong modulations due to the exponentially growing perturbation.
	Furthermore, the initially constant mean-flow horizontal wind experiences acceleration.
	The analytic maximum instability growth rate according to \eqref{eq:limrela} is 2.2 for this particular case.
	In terms of an approximated $L_2$-norm that we compute numerically, the actual growth rate of the Gaussian perturbation is found to be 1.6\,.
 	The difference between theoretical and observed rate occurs because the perturbation is not optimal.


	\section{Summary and discussion}
	\label{sec:conclusion}

	In this paper we investigated nonlinear mountain waves
	which are governed by Grimshaw's dissipative modulation equations
	being asymptotically consistent with the compressible Navier-Stokes equations
	and the dissipative pseudo-incompressible equations likewise.

	We introduced a wave-Reynolds number characterizing the stationary solution.
	When this dimensionless quantity is of order unity, the wave amplitude saturates by small-scale instabilities 
	and assumes a maximum as anelastic amplification by the thinning background air is exactly balanced by the turbulent damping.
	We analyzed this regime with respect to modulational stability as nonlinearities dominate for large-amplitude waves.
	It turned out that transient instabilities emerge that propagate upwards.
	We tested this analysis solving the modulation equations numerically and found excellent agreement with the theory.

	In the framework of saturated nonlinear wave theory, that we presented in this paper,
	the saturated mountain wave does initially not accelerate the mean-flow horizontal wind.
	Instead, a mean-flow horizontal kinematic pressure gradient emerges that keeps the horizontal wind constant
	by balancing the viscous forces. 
	The wave persists structurally and loses energy directly to turbulence 
	which in turn damps altitudinal amplification.
	Eventually, the mean flow is accelerated by a transient instability in the saturation zone propagating upwards while growing 
	and transferring kinetic energy to the mean flow.
	
	Our investigations have two implications being of interest for gravity wave parametrizations 
	in numerical weather prediction and climate modeling.
	First, the induced mean-flow behaves wave-like. 
	Its evolution is governed by a dispersion relation for the linear perturbation.
	In conclusion, it may be interpreted as an upward-traveling secondary wave of larger scale than the primary wave.
	Secondary waves with wavelengths comparable to the primary modulation scale were investigated by \citet{Vadas2002,Vadas2003,Becker2018a}.
	The authors propose a generating mechanism based on body forces produced by the dissipating primary wave.
	In contrast to this model, our secondary wave is generated by direct 
	wave-mean-flow interaction that compares to \citet{Wilhelm2018}.

	Second, this novel picture of saturated mountain waves may explain the bias 
	between the onset of small-scale instability and the actually observed mean-flow acceleration \citep{Achatz2007}.
	We give an extension to the established picture where waves become unstable at the breaking height
	and deposit their momentum and energy onto the mean flow at this level. 
	As it turns out, only in combination with modulational instability does an initially saturated nonlinear wave induce a mean flow.
	This modulational instability has its own transient velocity and growth rate 
	which we can compute explicitly.
	Then, the mean-flow acceleration depends on these two quantities.

	In our derivations we assumed piecewise analytic solutions being matched.
	This assumption can be weakened to almost arbitrary background, fulfilling hydrostatics and the ideal gas law,
	as well as almost any mean-flow horizontal wind.
	However, these functions of height must be restricted to converge to constant values at the infinities.
	The resulting spectrum describing the temporal evolution of the perturbation would be much more complicated 
	but still analytically assessable by Fredholm operator theory \citep{Schlutow2018}. 
	In conclusion, our results are valid in a much more realistic atmosphere.

	\section*{Acknowledgments}
		The author thanks Prof. Ulrich Achatz and Prof. Rupert Klein for
	many helpful discussions during the preparation of this article.
	The research was supported by the German Research Foundation (DFG) through grants KL 611/25-2 of the
	Research Unit FOR1898 and Research Fellowship SCHL 2195/1-1.

	\appendix	
	
	\section{Well-posedness: Are the instability growth rates finite?}
	\label{app:weelpos}

	In this appendix we demonstrate that the linear differential operator of the EVP \eqref{eq:evp} is well-posed.
	This property is essential for existence, 
	uniquenes and continuous dependency of the solution for the perturbation on the initial data.
	In section \ref{sec:stability} we showed that the spectrum of the operator extents 
	into the right half of the complex plane rendering the saturated wave unconditionally unstable. 
	But is the spectrum bounded from above on the real axis?
	Let us denote the discriminant of the square root appearing in $\lambda_{1,2}$ of \eqref{eq:spectrum} by $D_Y(\mu)$.
	The real part being the instability growth rate may be expressed by
	\begin{align}
		\Re(\lambda_{1,2}(\mu))=\pm K_x\sqrt{A}\,|D_Y(\mu)|^{1/2}\cos\left(\frac{1}{2}\arg(D_Y(\mu))\right)
	\end{align}
	where
	\begin{align}
		|D_Y(\mu)|^{1/2}&=|\mu|\sqrt[4]{(2\Lambda K_z-H^{-1}\Omega'')^2\mu^{-2}+\Omega''^2}\\
		\arg(D_Y(\mu))&=\begin{cases}
			\arctan\left(\frac{H^{-1}\Omega''-2\Lambda K_z}{\Omega''\mu}\right), &\Omega''<0\\
			\arctan\left(\frac{H^{-1}\Omega''-2\Lambda K_z}{\Omega''\mu}\right)\pm\pi, &\Omega''>0.
		\end{cases}
	\end{align}
	The function $\Re(\lambda_{1,2}(\mu))$ has no poles. Thus, we are only concerned with its behavior at the infinities.
	We have
	\begin{align}
		|D_Y(\mu)|^{1/2}&=\mathit{O}(|\mu|)\quad\text{as }\mu\rightarrow\pm\infty,\\
		\cos\left(\frac{1}{2}\arg(D_Y(\mu))\right)&=
		\begin{cases}
			\mathit{O}(1), &\Omega''<0\\
			\pm\frac{1}{2}\frac{H^{-1}\Omega''+2\Lambda K_z}{\Omega''}\mu^{-1}+\mathit{O}(|\mu|^{-3}), &\Omega''>0
		\end{cases}\quad\text{as }\mu\rightarrow\pm\infty.		
	\end{align}
	Therefore,
	\begin{align}
		\label{eq:limrela}
		\Re(\lambda_{1,2}(\mu))\rightarrow\begin{cases}
			\pm\infty, &\Omega''<0\\
			\pm\frac{1}{2}\frac{2\Lambda K_z-H^{-1}\Omega''}{\sqrt{\Omega''}}, &\Omega''>0
		\end{cases}\quad\text{as }\mu\rightarrow\pm\infty.
	\end{align}
	In conclusion the operator is well-posed \citep{Kapitula2013} if $\Omega''>0$ and ill-posed otherwise.



	\bibliographystyle{plainnat}
	\bibliography{library.bib}

\end{document}